\documentstyle[psfig]{l-aa}
\begin{document}
\thesaurus{03(11.01.2, 11.05.2, 11.09.2, 11.12.2, 13.09.1, 13.19.1)}

\title {Cosmologically distant OH megamasers: A test of the galaxy
merging rate at $Z \approx 2$ and a contaminant of blind HI surveys
in the 21cm line}

\author{F. H. Briggs }
\institute{Kapteyn Astronomical Institute, P.O. Box 800, 9700 AV
  Groningen, The Netherlands 
}

\offprints{fbriggs@astro.rug.nl} 
\date{Received 30 September 1997; accepted 20 May 1998} 

\maketitle

\markboth{F.H. Briggs:  OH MegaMasers}{ }

\begin{abstract}  

Bright OH megamaser galaxies, radiating 1667/1665 MHz lines, 
could be detected at redshifts from $z \approx 1$ to 3 in moderate
integration times with existing radio telescopes. The  
superluminous FIR galaxies that host the megamasers
are relatively rare at $z\approx 0$, but they may have
been more common at high redshift, if the galaxy merger rate increases
steeply with redshift.  Therefore, blind radio spectroscopic surveys at
frequencies of 400 to 1000 Mhz can form an independent test of the
galaxy merger rate as a function of time over the redshift interval
$z =$ 4 to 0.7.

The redshift range $z = 0.17$ to ${\sim}0.4$ will be difficult to
survey for OH masers, since spectroscopic survey signals will be confused
with HI emission from normal galaxies at redshifts less than 0.3.
In fact, the
signals from OH masers are likely to dominate over 21cm line emission from
normal galaxies at frequencies
below 1200 MHz (i.e. large redshifts $z_{HI} > 0.18 $ and $z_{OH} > 0.4$).
Surveyors  of nearby galaxies in the 21cm line may find that
OH masers form a contaminant to
deep, blind HI surveys for redshift
velocities less than a few hundred kilometers per second.  
At frequencies just above
1420 MHz, sensitive sky surveys might detect
OH masers, which could be mistaken for a population of ``infalling,
compact High Velocity Clouds'' but would ultimately be traced to
luminous FIR background galaxies at $z \approx 0.17$ once optical and
IR follow-up has been performed.

\end {abstract}


\section{Introduction}

The detectability of OH megamaser and gigamaser galaxies at cosmological
distances has been noted by several authors  (Baan 1989, Baan et al
1992a, Norman \& Braun 1996, Baan 1997).
So far, the sample of known megamasers has been assembled from targeted
observations of luminous galaxies for which redshifts are known
(cf. Baan et al 1992b, Martin et al 1989, Stavely-Smith et al 1992).
The receiving systems are tuned to the predicted frequencies of
the redshifted OH lines whose rest frequencies fall at 1665 and 1667 MHz.
The greatest success rate for discovery of megamasers
occurs in galaxy samples that are selected
for having the highest far infrared (FIR) 60~${\mu}m$ luminosity, with
the result that ${\sim}50$\% of galaxies with 
$L_{60\mu}> 10^{11.2}h^{-2}L_{\odot}$ are strong OH masers
($h = H_o/(100{\rm\, km\,s}^{-1}$).  The odds decline
to ${\sim}5$\% for samples where  $L_{60\mu} \approx 10^{9}L_{\odot}$.
There is a striking correlation between OH luminosity $L_{OH}$ 
and $L_{60\mu}$, with  $L_{OH} \propto L_{60\mu}^2$ (Baan et al 1992a, 
Baan 1989, Martin et al 1989).  

A picture 
explaining both the probability of detection and the relative
strengths of megamaser emission from objects of different 
$L_{60\mu}$ has the maser strength depending on strength of
the starburst activity in the galaxy (Baan 1989, Henkel et al 1991).
Major merger and interaction events are the strongest stimulants of
star formation and also produce the most turbulent interstellar
gas distributions; these conditions provide the strongest IR fluxes
for pumping the maser levels, as well as the largest covering factors
of the star forming regions, leading to the greatest probability that
views from random directions will see OH maser activity.  Galaxies
with milder star formation have more placid disks, and maser emission
is then observed only by observers who view the galaxies nearly 
edge-on to their planes.  The strong association between the
ultra-luminous FIR galaxies and strongly-interacting and merging
systems has been recently discussed by Clements et al (1996).

At the present epoch, the superluminous FIR galaxies that host the
most luminous masers are relatively rare galaxies.  This means that
radio spectroscopic surveys of the sky would need to cover large
solid angles before they would begin to detect OH masers at random
from the local galaxy population.
On the other hand, this paper will show that, since OH emission can be
so very strong, it becomes probable that surveys with
large spectral bandwidths and covering large depths are likely to 
identify cosmologically distant megamasers with moderate integration
times.

The most luminous infrared sources known are the objects detected
at high redshifts, $z\approx 0.4$ to 4.7,
(Rowan-Robinson 1996, and references therein) with
$L_{FIR} \sim 10^{12}$ to 10$^{14}h^{-2}L_{\odot}$. There is general
consensus that the merging and interaction rate of galaxies was
greater in the past with merging rate rising in proportion
to $(1+z)^m$. Values for
$m$ as high as 4  (Carlberg 1992, Lavery et al 1996) 
are mentioned in the literature, although
there is some observational evidence pointing to $m\approx 1.2$ for
$z<1$ in the HST Medium Deep Survey (Neuschaefer et al 1997)
and $m=2.8\pm0.9$ from faint galaxy catalogs compiled from CFHT
surveys (Patton et al 1997).
The comoving number density of bright QSOs ($M_B<-24.5$ $+5\log(h)$)
increases as $(1+z)^{6}$ from $z=0$ to 2 (Hewett et al 1993), and, if the
QSO phenomenon is tied to galaxy interactions, this steep evolution
could point to a steep rise in the interaction rate. 

Surveys in the range 400 to 1000 MHz may provide an independent
measure of the merging rate by determining the density of merging
galaxies as a function of redshift.  Since some of the currently
most viable models of galaxy formation require the bulk of the
construction of the larger galaxies to occur by merging throughout
this  range of redshift ($z \approx 0.5$ to 3), 
surveys for OH masers will form an
important test of the principal processes of galaxy formation.

\section{Detection rate of cosmologically distant OH Masers}

The detection rate for OH megamasers can be estimated by combining 
1) knowledge of the density of prospective FIR luminous host galaxies,
2) the probability that galaxies of each luminosity will be seen as
megamaser emitters, 
3) the dependence of megamaser strength on FIR
luminosity, and 
4) the sensitivity of radio telescope receiving
systems.  The steps in the estimation of the detection rate
are outlined in the following subsections. A parallel line of
reasoning applies to the detectability of normal galaxies in the
21cm line of neutral hydrogen, as discussed in Sect. 2.5.

\subsection{The FIR luminosity function $\Phi(L_{60\mu})$}

Koranyi \& Strauss (1997) have presented a conveniently formulated
description of the far-infraed luminosity function, based on complete
samples of galaxies from the IRAS 60~$\mu$m survey.  Their paper 
analyzes the sample in the context of several cosmological models;
here, parameters for the ``$p=1$'' case, which is the conventional
cosmology, are adopted to describe the number density of luminous,
FIR-selected galaxies.

The 60~$\mu$m differential luminosity function is given by

\begin{equation}
\Phi(L_{60}) = \left( \frac{\alpha}{L_{60}}+\frac{\beta}{L_{60}^*+L_{60}}\right) \Psi(L_{60})
\end{equation}
 with $\Psi(L_{60})$, the cumulative luminosity function, defined as
\begin{equation}
\Psi(L_{60}) = C\left( \frac{L_{60}}{L_{60}^*}\right)^{-\alpha}
          \left(1 + \frac{L_{60}}{L_{60}^*}\right)^{-\beta}.
\end{equation}
For the case of conventional cosmology, Koranyi \& Strauss fit values
of $\alpha= 0.49$, $\beta = 1.81$ and $L_{60}^*=10^{9.68}L_{\odot}$.  
The normalization constant, $C$,
is related to their parameter $n_1 = 0.058$~Mpc$^{-3}$,
through the relation $n_1 = \Psi[L_{min}(z_s)]$, where $L_{min}(z_s)$
is the minimum luminosity detectable in the sample, whose members
are restricted
to lie at redshifts greater than $z_s$.  For Koranyi \& Strauss' analysis, 
$L_{min}(z_s)/L^* = 8.8{\times}10^{-3}$, which leads to $C=5.8{\times}10^{-3}$
Mpc$^{-3}$ ($H_o = 100$~km~s$^{-1}$~Mpc$^{-1}$).

In the $L_{60}>>L_{60}^*$ regime, which will characterize the hosts of the
brightest OH line emitters, $\Phi(L_{60})$ may be approximated as
\begin{equation} \label{phiapprox.eq}
\Phi(L_{60}) \approx 
  C(\alpha+\beta)L_{60}^{*-1}(L_{60}/L_{60}^*)^{-(\alpha+\beta+1)}.
\end{equation}

\subsection{The effective OH luminosity function $\Theta(L_{OH})$}

\begin{figure}
\psfig{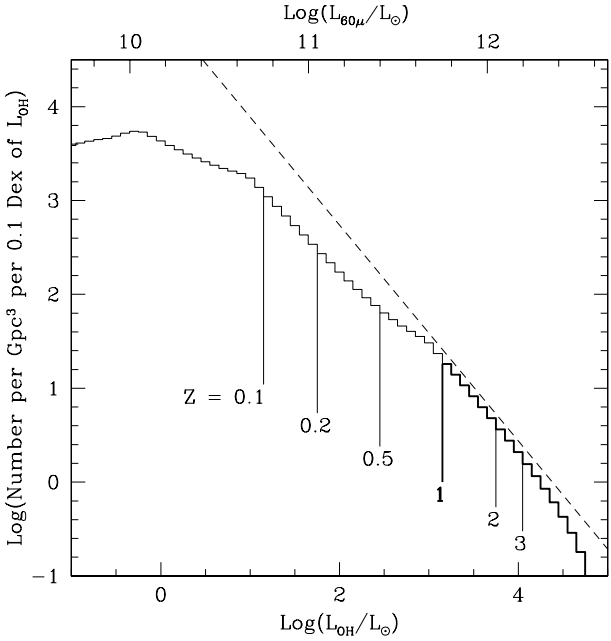}
  \caption[]{Estimate for the effective OH megamaser luminosity function
   at $z\approx 0$. The vertical axis gives the apparent number of
   observed per Gpc$^3$ between $10^{-0.05}L_{OH}$ and $10^{+0.05}L_{OH}$,
  with $\log(L_{OH})$ in solar luminosities plotted along the horizontal
  axis. The approximate FIR luminosity, $L_{60}$ of these galaxies is 
  labeled along the top border. With a observational detection limit
  of 1~mJy in the OH lines, only a portion of the high luminosity 
  side of the function would  be detectable at high redshifts, as indicated
  by the labeled cut-off luminosities.  The  portion detectable at $z=1$
  is drawn as a bold line.($H_o = 100$~km~s$^{-1}$~Mpc$^{-1}$). The dashed
  line is the approximation for $\hat\Theta(L_{OH})$ from 
  Eq.(\ref{approxlmfnct.eqn}) with $f_m=1.0$.}
  \label{OHLumfnct.fig}
\end{figure}

The strong correlation between OH and 60$\mu$m luminosity, 
$L_{OH}\propto L_{60}^2$
(cf. Baan et al, 1992a) 
provides a way to recast the 60$\mu$m luminosity function for
luminous FIR galaxies to obtain 
an OH megamaser luminosity function $\Theta(L_{OH})$ as shown in 
Fig.~\ref{OHLumfnct.fig}.
In this paper, this has been accomplished in two complementary ways:
one by numerical integration of $\Phi(L_{60})$ to  obtain
the number of OH sources falling in logarithmically spaced bins of $L_{OH}$,
and the second by noting that the masers that  are detectable at
cosmological distances inhabit host galaxies that are far more luminous
than $L_{60}^*$ and then using the approximation in Eq.(\ref{phiapprox.eq})
to derive an analytic expression.  

The advantages of computing $\Theta(L_{OH})$ numerically are that it
(1) simplifies inclusion of the spread of OH luminosities in a band
of width ${\sim}10^{{\pm}0.7}$ about the $L_{OH}\propto L_{60}^2$
relation (Baan et al 1992a), and (2) allows for varying the probability $f_m$
that luminous FIR galaxies will appear to be megamasers as a function
of $L_{60}$.  Here, the range of FIR luminosities that are considered
to contribute OH megamasers is conservatively restricted to the range
where strong OH emission has actually been detected (cf. Fig. 2 of
Baan et al 1992a); adopting $H_o=100$~km~s$^{-1}$Mpc$^{-1}$, this
range is $5{\times}10^9<L_{60}<2{\times}10^{12}L_{\odot}$. Within this
range, $f_m$ takes on non-zero values as follows, 
$f_m=0.05$ for $5{\times}10^9<L_{60}<2{\times}10^{10}L_{\odot}$,
$f_m=0.2$ for $2{\times}10^{10}<L_{60}<2{\times}10^{11}L_{\odot}$,
and
$f_m=0.5$ for $2{\times}10^{11}<L_{60}<2{\times}10^{12}L_{\odot}$
(from Baan 1989, Baan et al 1992b).
Since these probabilities are likely to be a result of viewing angle,
the derived luminosity function should be considered as an
``effective OH luminosity function'' that provides the probability
of detection for randomly oriented sources, rather than an 
an accurate accounting of the number of OH emitting galaxies.
The constant of proportionality $C_1$ in $L_{OH}= C_1 L_{60}^2$
is $4.5{\times}10^{-21}L_{\odot}^{-1}$, and a fiducial $L_{OH}^*$
can be defined to be  $L_{OH}^*= C_1L_{60}^{*2}= 0.10L_{\odot}$.

Figure~\ref{OHLumfnct.fig}
shows the numerically computed luminosity function for OH
megamasers in logarithmically spaced bins of 0.1 Dex.  For reference,
the 60~$\mu$m luminosity ($L_{60}=\sqrt{L_{OH}/C_1}$)
is indicated on the top axis.  Since the
bright OH megamasers that are likely to be useful in cosmological
studies lie well above $L_{OH}^*$, the approximation in Eq.(\ref{phiapprox.eq})
can be used to verify the numerical estimate in the range of greatest interest:
\begin{eqnarray}
\label{approxtheta.eqn} 
\Theta(L_{OH})dL_{OH} &=& 
f_{m}\Phi[L_{60}(L_{OH})]\left(\frac{dL_{60}}{dL_{OH}}\right)dL_{OH}
\\
 &=& Cf_{m}\frac{(\alpha+\beta)}{2}
  \left(\frac{L_{OH}}{L^*_{OH}}\right)^{-\frac{\alpha+\beta+2}{2}} 
  \frac{dL_{OH}}{L^*_{OH}}
\nonumber
\end{eqnarray}
Since it is convenient to plot the number of objects
per logarithmically spaced bins of luminosity, the number of objects
falling in bins of
$\Delta L_{OH}=0.1$ Dex centered on $L_{OH}$ can be obtained from
\begin{eqnarray}
\hat\Theta(L_{OH}) &=& 
\Theta(L_{OH})\Delta L_{OH} 
\nonumber\\
 &=& \Theta(L_{OH})(10^{0.05}-10^{-0.05}) L_{OH}
\nonumber\\
 &=& 0.231 Cf_{m}\frac{(\alpha+\beta)}{2}
\left(\frac{L_{OH}}{L^*_{OH}}\right)^{-\frac{\alpha+\beta}{2}}
\label{approxlmfnct.eqn} 
\end{eqnarray}
This function, with $f_m$ set equal to 1, is plotted in 
Fig.~\ref{OHLumfnct.fig} for comparison with the result of the
numerical integration.

In a related analysis,
Baan (1997) presents the 60${\mu}$ FIR luminosity function for OH
megamaser sources as a function of $L_{60\mu}$ for direct comparison with 
the luminosity function for broader population of all FIR galaxies.
Baan (1997) further bins the fraction $f_m$ in 0.5 Dex rather than the
1~Dex divisions used in the approximation here. The finer binning
would lead to less abrupt inflections in the OH luminosity function
derived here (cf. Fig.~\ref{OHLumfnct.fig}).  Overall, there is good
agreement between the spatial densities of galaxies between Baan (1997)
and the present work, although the fraction of OH galaxies showing
megamaser emission presented by Baan (1997) appears to be roughly a 
factor of two lower in each luminosity range 
than in the earlier publications.  This factor of two
uncertainty can be propagated
through the subsequent estimates for volume densities and densities per
solid angle.

\subsection{Profile width and detectability at large redshift}

\begin{figure}
  \psfig{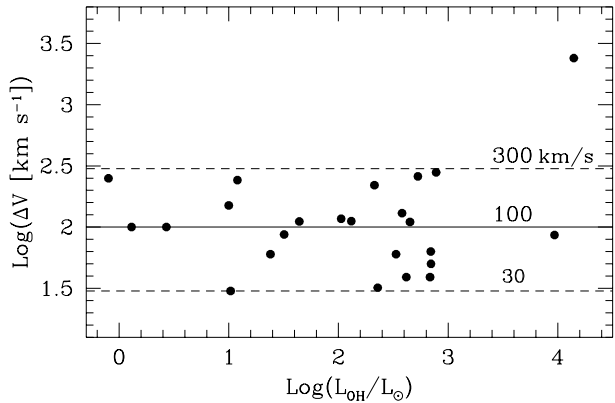}
  \caption[]{Velocity widths of OH megamaser profiles as a function of
  luminosity in the 1667 OH line. Data from Staveley-Smith et al 1992,
  Martin et al 1989, Baan et al 1992a, Baan et al 1992b. }
  \label{velwid.fig}
\end{figure}

The integral line flux observed from a cosmologically distant emitter at 
redshift $z$ is (Wieringa et al 1992)
\begin{eqnarray}
S &=& \frac{L_{OH}}{4\pi d_L^2}
\label{cosmflux.eqn}\\
&=& \frac{L_{OH}}{\pi}\left(\frac{H_o}{4c}\right)^2
\frac{\Omega_o^4}
     {[\Omega_oz+(\Omega_o-2)(\sqrt{1+\Omega_oz}-1)]^2}\nonumber
\end{eqnarray}
where $d_L$ is the luminosity distance, 
$H_o$ is the Hubble constant and $\Omega_o$ is the 
cosmological mass density normalized to the critical density.
If the line of width $\Delta V$~km~s$^{-1}$ is 
emitted at rest frequency $\nu_o$ and redshifted
to $\nu= \nu_o/(1+z)$, the observed frequency width of the line will  be 
$\Delta\nu \approx  \nu\,\Delta V/c$, providing an average
flux density over the line profile of
$S_{\nu}\approx S/\Delta\nu\propto S(1+z)$.  Thus, 
the line width is squeezed in frequency as a line of velocity
width $\Delta V$ is observed toward higher redshifts, leading 
to a partial compensation for the reduction in the integral line
flux by the inverse square law (Eq.(\ref{cosmflux.eqn})),
The range of line widths observed for
OH megamaser galaxies is illustrated in Fig.~\ref{velwid.fig}, where
the line width is plotted as a function of maser luminosity.  The vast
majority of the line profiles fall in the range 30 to 300 km~s$^{-1}$.
Only the most luminous source known shows a much broader profile
(Baan et al, 1992a).

The sensitivity of a radio telescope is generally specified (cf. Crane \&
Napier 1989) in
terms of its receiver noise level as parameterized by a system 
temperature, $T_{sys}$, and the antenna sensitivity, 
$K_a = 10^{-26}\epsilon_{ap}A/2k$, which is a relation between
flux density, $S_{\nu}$ in Jy and antenna temperature $T_a=K_a S_{\nu}$ K.
The quantities $\epsilon_{ap}$, $A$, and $k$ are the aperture efficiency,
the antenna collecting area and the Boltzmann constant in mks units,
respectively.
The noise level attained by observing a single polarization 
with a spectral resolution of $\Delta\nu$ and an integration
time $\tau$ is   $\sigma_{Jy}=T_{sys}K_c/(K_a\sqrt{\Delta\nu\,\tau})$
where $K_c$ is a factor with values typically between 1 and 1.4 depending
on the details of the digital correlation spectrometer used to obtain
the spectra. 

\begin{figure}
  \psfig{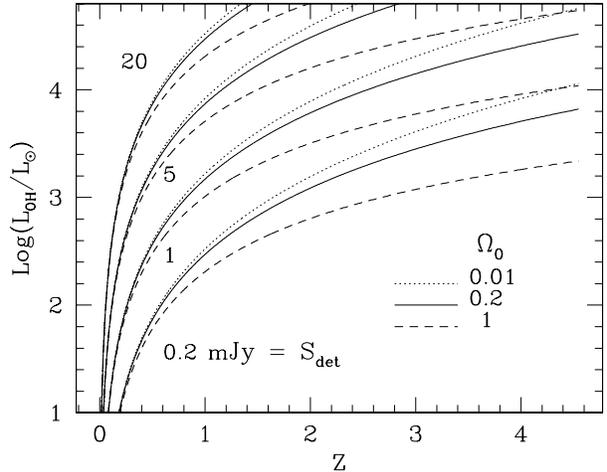}
  \caption[]{Minimum $L_{OH}$ detectable at redshift $z$ in an
   observation with detection threshold $S_{det}$.  Families of
   curves are drawn for $S_{det} =$ 0.2, 1, 5, and 20 mJy 
   for cosmological models with $\Omega_o =$ 0.01, 0.2 and 1.
   Models assume $H_o = 100$~km~s$^{-1}$~Mpc$^{-1}$ and OH profiles
   having width 100~km~s$^{-1}$.}
  \label{deteclev.fig}
\end{figure}

As an example of sensitivities that can be attained at present or in the
near future, consider the specifications and noise levels for typical
observations with the Westerbork Synthesis Radio Telescope. For this
system, receivers at 600 and 1400 MHz will have $T_{sys}$ of 45 and 25~K
respectively.  The effective $K_a$ of combining fourteen 25~m telescopes
is $K_a\approx 1.4$ K~Jy$^{-1}$.  For velocity widths of 100 km~s$^{-1}$,
these receiving systems will reach the 1~mJy ($5\sigma$) level in roughly
one to three 12 hour long integrations. The WSRT telescope will soon have
the further benefit of a spectrometer capable of simultaneous observation
in the synthesis mapping mode
of an 80 MHz bandwidth at frequencies below 1200 MHz and 160 MHz above
1200 MHz with adequate spectral
resolution for the identification of 100 km~s$^{-1}$
width signals.
The minimum detectable OH line luminosity as a function of redshift is
plotted in Fig.~\ref{deteclev.fig} for a range of cosmological
models and observational sensitivity.

\subsection{Density of megamasers per solid angle}

\begin{figure}
  \psfig{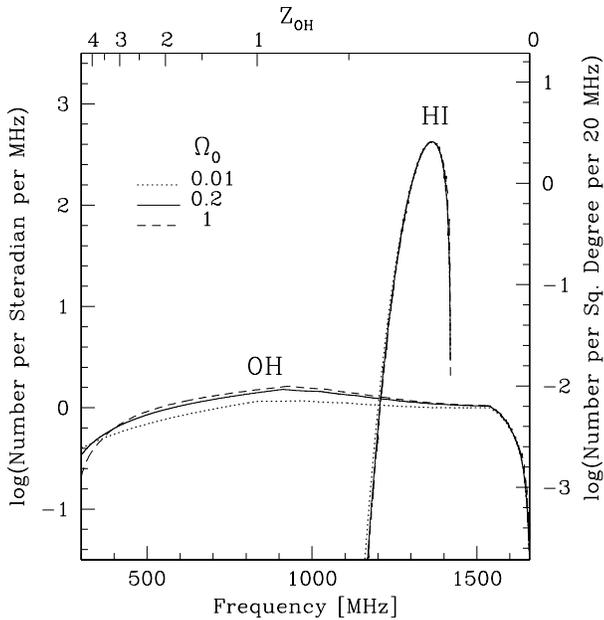}
  \caption[]{Detection rate of signals from OH megamasers and HI in
   normal galaxies as a function of frequency. Computed for line widths
  of 100 km~s$^{-1}$ and survey sensitivity of 1~mJy. The OH redshift
  for the OH lines is indicated at the top of the figure.
  The relatively narrow peak in detection rate for neutral hydrogen
  is labeled ``HI,'' while ``OH'' refers
  to the broader, low-level function of redshift. Three cosmological
  models are shown: $\Omega_o =$ 0.01, 0.2 and 1. }
  \label{cosmcompare.fig}
\end{figure}

Estimates for the number of detectable megamasers in a blind survey
of a large area of sky can be straightforwardly obtained by
computing the comoving volume at redshift $z$ contained in the 
solid angle $d\Omega$  and depth $dz$ (Wieringa et al 1992),
\begin{eqnarray}
dV &=& 4\left(\frac{c}{H_o}\right)^3
\frac{[\Omega_oz+(\Omega_o-2)(\sqrt{1+\Omega_oz}-1)]^2}
     {\Omega_o^4(1+z)^3\sqrt{1+\Omega_oz}}  dz\,d\Omega, \nonumber
\end{eqnarray}
and then filling the volume with OH emitters according to the density
prescribed by Fig.~\ref{OHLumfnct.fig}.
A count of the number of objects whose observed flux exceeds
a detection threshold can be made and plotted as shown for three
cosmological models in Fig.~\ref{cosmcompare.fig}.  The detection
level was chosen to be 1 mJy for 100 km~s$^{-1}$, consistent with
feasible integration times with existing radio telescopes.
In Fig.~\ref{cosmcompare.fig}, 
the horizontal axis is chosen to be observed frequency
and the counts are binned according to expected detection rate
per bandwidth  of radio
spectrum.  These are natural units for use in planning surveys.
In computing this diagram, the assumption was made that
there is no evolution in the comoving
number densities or luminosities of the OH megamasers with redshift.
Curves for showing the detection rate for the confusing signals 
from normal galaxies emitting the 21cm hydrogen line are shown for
comparison. Computations of the detection rate for a range of
sensitivities is given in  Fig.~\ref{ratecombo.fig}.

\subsection{Confusion with 21cm line emission from normal galaxies}

The expected number of detections of normal galaxies in the 21cm line
of neutral hydrogen can be computed in a similar manner to that for
OH megamasers.  The HI luminosity function $\phi(M_{HI})$
observed for nearby galaxies (Zwaan et al 1997) 
can be represented as a function of
HI mass, $M_{HI}$
\begin{equation}
\phi(M_{HI})\,  dM_{HI} = \phi^*\left(M_{HI}/M_{HI}^*\right)^{-\gamma}
   e^{- M_{HI}/M_{HI}^*}\,dM_{HI}
  \label{himassfnct.eqn}
\end{equation}
with $\phi^*=0.014$~Mpc$^{-3}$, $\gamma=1.2$ and 
$M_{HI}^* = 10^{9.55}M_{\odot}$. The luminosity in the 21cm line
from an optically thin
cloud of neutral hydrogen with  HI mass $M_{HI}$ is
$L_{HI}/L_{\odot}=6.2{\times}10^{-9}M_{HI}/M_{\odot}$.

Curves for detection rates in the hydrogen line are plotted in 
Figs.~\ref{cosmcompare.fig} and \ref{ratecombo.fig}
for comparison with the OH line strengths.  The exponential cutoff
to the HI luminosity function causes the HI detection rate to cutoff
very hard around $z=0.2$ for 1~mJy sensitivity.  The power law
function for the OH masers permits the bright end of the OH
maser population to dominate the detection rate at frequencies below
${\sim}$1200 MHz for reasonable integration times with existing 
telescopes.

Not only are the typical velocity widths of the OH profiles comparable
to velocity spreads measured in the 21 cm line for low mass and face-on 
galaxies, but also the pair of strong OH lines at 1665.4 and 1667.4 MHz
could produce a spectrum that, at low signal to noise ratio, could
imitate a double horned galaxy profile with a splitting of ${\sim}$360
km~s$^{-1}$, although the horns are likely to be asymmetric due to the
usual weakness of the 1665 line relative to the 1667 line.

Sifting the OH masers from the HI lines from normal galaxies of low
redshift would
be straightforward.  High spatial resolution radio observations would
show OH masers to have high brightness temperature and would
identify the sources with compact optical sources for which emission
lines would provide an unambiguous redshift.   OH megamasers are
strong emitters of 60~$\mu$m infrared,  but sources at $z\approx 0.17$
(or greater), selected for OH line strength around 1~mJy, will have
 60~$\mu$m flux densities approximately equal or less than the
0.2 Jy sensitivity level of the IRAS 60~$\mu$m Faint Source Survey
(Moshir et al 1992). Since the infrared spectrum of the OH megamasers
falls steeply from 60~$\mu$m to 25~$\mu$m, suitable megamaser host
galaxies at moderate to
high redshift may become increasingly difficult to detect at
infrared wavelengths as the 60~$\mu$m flux is redshifted toward
still longer wavelengths.

\section{Redshift dependent merging rate}

\begin{figure}
  \psfig{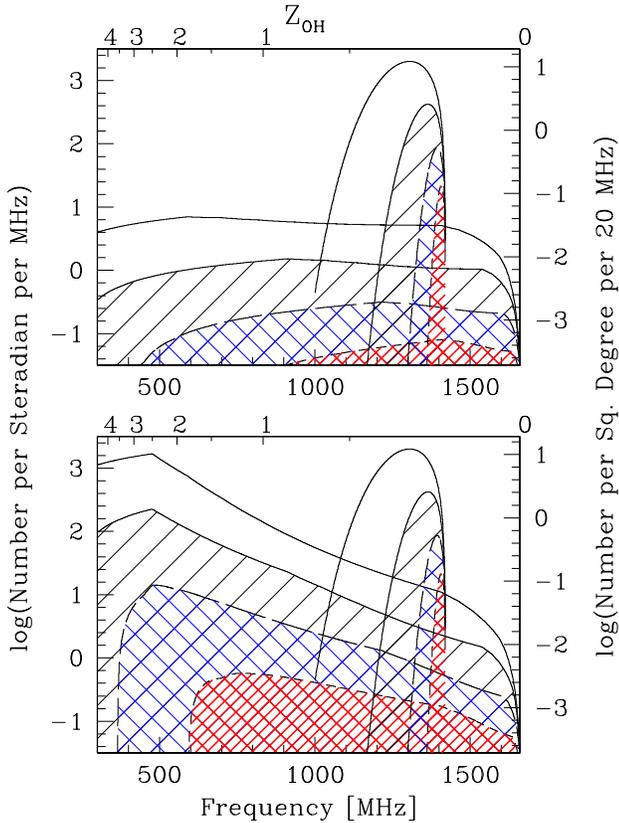}
  \caption[]{Detection rate as a function of frequency
   for two different redshift dependences of the merging rate.  
  {\it Top Panel} ~ Constant merging rate. {\it Bottom Panel} ~
  Merging rate proportional to $(1+z)^{4.5}$ for $z<2.5$ but 
  constant for $z>2.5$.  Detection rate is
   computed in a cosmological model with $\Omega_o = 0.2$
   for four sensitivity levels (0.2, 1, 5, and 20 mJy) for
   signals of velocity width 100~km~s$^{-1}$. The lowest sensitivity
   survey (20 mJy) corresponds to the most heavily shaded area, and the highest
   sensitivity (0.2 mJy) is unshaded.
   }
  \label{ratecombo.fig}
\end{figure}

The  importance of merging small galaxies to build the large galaxies
that we observe at the present is a topic of considerable controversy,
since the question addresses the foundations for theories 
of galaxy formation.  In this section,
it is demonstrated that surveys sensitive to OH masers at $z\approx 2$
may be able to measure the number density of interacting and merging
systems at high redshift at a time when the merging rate is thought
to peak. Observations using radio telescopes with sufficient sensitivity
and sufficient spectral bandwidth can not only determine the statistics
of the evolving density of OH masers with time, but they will also
select systems of interest at high redshift, which can be further
studied with a variety of techniques to   determine physical
conditions, dynamics and effects of local environment.

For the purposes of estimating the effect of an increased merging rate 
at early times, the model explored in 
Fig.~\ref{cosmcompare.fig} and the top panel of Fig.~\ref{ratecombo.fig}
can be 
modified by increasing the number density of OH megamasers in proportion
to $(1+z)^m$, in keeping with the idea that, locally, OH megamasers
are associated with luminous FIR galaxies, which in turn are found
to occur in systems with merging and interacting morphologies (Clements
et al 1996).  For comparison, an extreme example, $m=4.5$ (Lavery
et al 1996) is shown in Fig.~\ref{ratecombo.fig}. The significance of
this detection rate is that an existing telescope, such as the 
Westerbork Synthesis Radio Telescope, which will reach a noise level
below 1 mJy 
level in a 12 hour synthesis observation and can survey a field
of one square degree per exposure at 600 MHz ($z_{oh}\approx 2$),
would be likely to detect a few objects per field, when it is equipped
with a spectrometer capable of 80 MHz (or more) bandwidth.

\begin{figure}
  \psfig{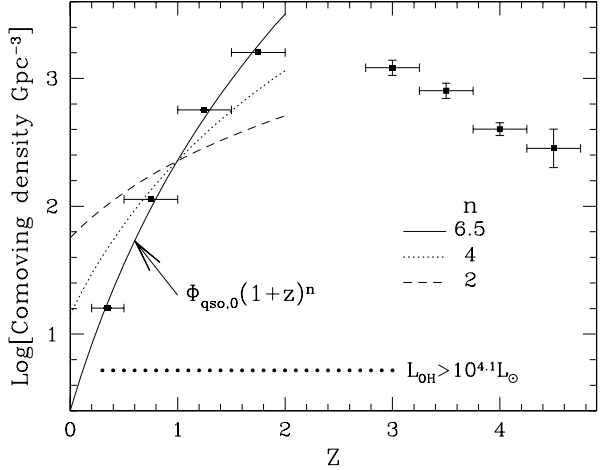}
  \caption[]{Comoving~density~of bright QSOs ($M_B<-24.5$ $+5\log(h)$) 
   as a function
   of redshift. Points at $z<2.5$ are derived from 
  Hewett et al (1993) (converted to $H_o=100$ km~s$^{-1}$ Mps$^{-1}$);
 points at $z>2.5$ are adapted from Schmidt et al (1995). 
   Three curves are drawn to indicate the evolutionary
   trend through the low redshift regime with the dependence
   $\propto (1+z)^n$ with $n=$ 2, 4 and 6.5.   A horizontal dotted line
   is drawn to indicate a constant comoving density equal to the
   current density (see Fig.~\ref{OHLumfnct.fig}) of luminous OH
   sources ($L_{OH}>10^{4.1}h^{-2}L_{\odot}$ and 
   $L_{60\mu}>10^{12.2}h^{-2}L_{\odot}$) that are bright enough to be
   detected at $z=3$ with 1~mJy sensitivity level.}
  \label{qsoevol.fig}
\end{figure}

The luminous quasar population shows a still stronger evolution 
than is typically suggested for normal galaxies. Since the most luminous
OH masers are associated with the most luminous galaxies in the
nearby universe, it is reasonable to speculate about a possible
relation between hosts for OH maser and quasars and the possibility
that both may be driven by mergers and interactions.
Figure~\ref{qsoevol.fig} shows recent observational constraints on
the comoving number density of bright QSOs ($M_B<-24.5$ $+5\log(h)$),
as presented by Hewett et al (1993) for $z< 2$ and Schmidt et al (1995) 
for $z>2.5$;  these studies provide evidence that the comoving
density peaked at $z\approx 2$ to 3.  
Representative curves for describing the increase in 
density $\phi_{-24.5}$ with redshift are drawn in the figure and indicate that
$\phi_{-24.5}$ is roughly proportional to $(1+z)^6$ to 
redshift as high as $z\approx 2$.  The QSO evolution 
is often described as an evolution in luminosity; Boyle et al (1988) quantify 
the QSO population as having a luminosity function 
$\phi_{qso}(L/L_*)\propto \phi_o(L/L_*)^{-3.7}$ with $L_*=L_o(1+z)^{3.2}$.
Integrating to obtain the number of objects per comoving volume
 brighter than $L_1$ gives 
\begin{eqnarray}
N(>L_1) &=& \int^\infty_{L_1/L_*}\phi_{qso}\left(\frac{L}{L_*}\right)
              d\left(\frac{L}{L_*}\right)
\\
&=& \frac{\phi_o}{2.7}\left(\frac{L_1}{L_*}\right)^{-2.7}
\nonumber\\
&=& \frac{\phi_o}{2.7}\left(\frac{L_1}{L_o}\right)^{-2.7}(1+z)^{5.9},
\nonumber
\end{eqnarray}
which has a proportionality that is consistent with the evolution for $z<2$
shown in Fig.~\ref{qsoevol.fig} and in Schmidt et al (1995).

The intention behind Fig.~\ref{qsoevol.fig} is to show that at least
one population of luminous object does evolve very steeply with
redshift. An association between AGN activity and galaxy merging
or interaction remains to be definitively established, although there
is evidence that ``warm,'' ultraluminous infrared galaxies may be
an evolutionary step in the evolution of optically selected
QSOs (cf. Surace et al 1998).
Interestingly, the $z\approx 0$ number density of ultraluminous
FIR galaxies is comparable to the $z\approx 0$ density of luminous QSOs,
as illustrated in the figure.
Furthermore, Saunders et al (1990) deduced a density evolution 
${\propto}(1+z)^{6.7}$ for their 60$\mu$m sample, although subsequent
study by Ashby et al (1996) did not find the tail to high redshift expected
if this were indeed the case. The evolution rate ${\propto}(1+z)^{4.5}$
explored in the lower panel of Fig.~\ref{ratecombo.fig} is fairly
mild in comparison to that of the QSOs.

\section{Contamination of blind HI 21cm line surveys}

\begin{figure}
  \psfig{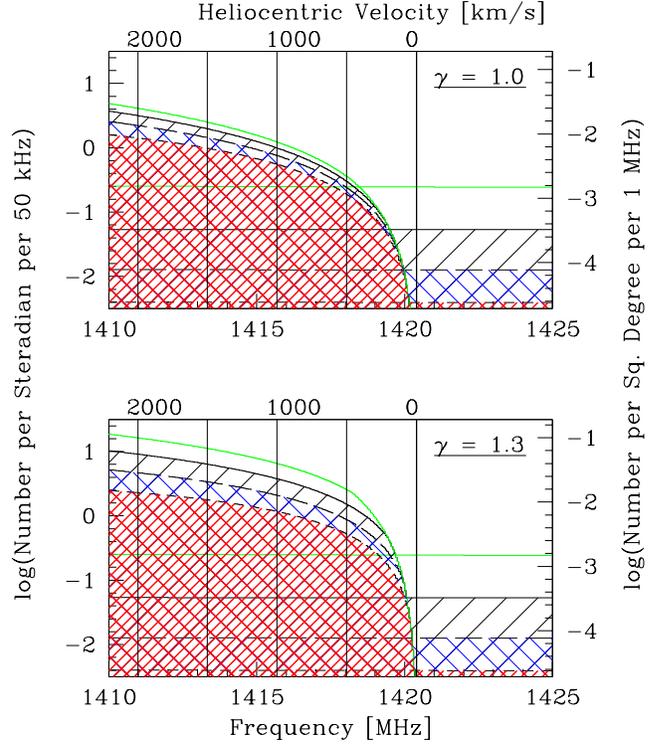}
  \caption[]{Detection rate of HI and OH signals near
  the rest frequency of neutral hydrogen. Two cases are illustrated for
  the faint end slope of the HI mass function, $\gamma=$ 1 and 1.3.
  Detection thresholds are set at 0.2, 1, 5 and 20 mJy.}
  \label{loZconfus.fig}
\end{figure}

Figures.~\ref{cosmcompare.fig} and \ref{ratecombo.fig}
 demonstrate that blind
spectroscopic surveys are more likely to detect distant OH megamasers
than HI from normal galaxies in the frequency bands below ${\sim}1200$MHz,
assuming that both megamasers and normal galaxies have similar properties
to those that are observed nearby. Above the rest frequency of the
hydrogen line at $\nu_{HI} =$ 1420.4 MHz and in a small range of 
frequencies just below $\nu_{HI}$ there is a possibility for confusing
OH emission at $z_{OH}\approx 1667/1420-1 = 0.17$ with weak HI signals 
from nearby dwarf galaxies and HI clouds.  To explore this possibility
in more detail, Fig.~\ref{loZconfus.fig} shows a
a plot similar to Fig.~\ref{ratecombo.fig} with an expanded scale
around 1420 MHz.  The vertical axes have been adjusted to more
convenient units to address this question.   The possibility for
confusion might arise in ``blind surveys'' for dwarf galaxies
in the HI line, where signals at noise levels of a few $\sigma$
are selected as candidate dwarfs. At these sensitivity levels, only
a narrow 1667 MHz might be detected, mimicing a narrow HI profile
of a few 10's of km~s$^{-1}$ velocity width. Clearly more sensitive
followup  in the radio line, as well as far infrared and
optical observations, would quickly show the distinction.

Some questions remain about the slope of the luminosity function for
galaxies at the faint end, and this uncertainty propagates into 
the faint galaxy tail of the HI-mass function as well. Recent HI surveys
have produced values for $\gamma$ (as defined in Eq.(\ref{himassfnct.eqn}))
 in the range 1.2 to 1.7 (Zwaan et al 1997,
Schneider 1997), although there is also observational evidence for
a sharp
decline in the number density of HI rich objects at faint HI luminosity
(Hoffman et al 1992), at
least in the Virgo cluster. Figure~\ref{loZconfus.fig} implies that
surveys with sensitivities of $\sim$1~mJy or better may be likely to
detect OH masers (at $z_{OH}\approx 0.17$) in comparable numbers to
the rate at which they detect weak HI signals in the velocity 
range $v_{HI}< 500$~km~s$^{-1}$.

\begin{figure}
  \psfig{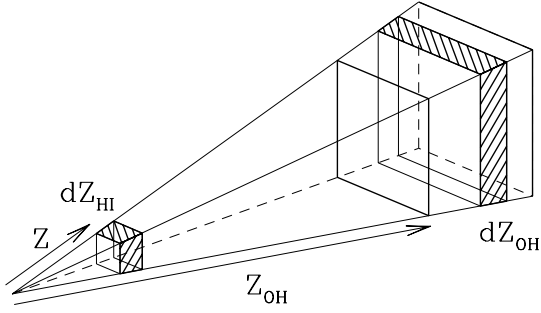}
  \caption[]{The volume $dV_{OH}\approx z_{OH}^2dz_{OH}d\Omega$ containing
 OH masers that is mapped into volume $dV_{HI}=z^2dz_{HI}d\Omega$ in HI
 surveys.}
  \label{sketch.fig}
\end{figure}

This regime  can be explored in more detail by estimating the form
of the HI mass function that the OH masers would imitate. 
Figure~\ref{sketch.fig} is an aid to visualize the problem. It shows
a volume $dV_{OH}$ at $z_{OH}$ that is mapped into the volume $dV_{HI}$
at $z$ due to the overlap 
that occurs in observed frequency for the two populations.
For simplicity in this estimate,
approximations for the luminosity distance $d_l\approx z_{OH}c/H_o$
and for the comoving volume $dV_{OH}\approx z_{OH}^2dz_{OH}d\Omega$
will be assumed, instead of the more rigorous forms used earlier
in this paper in the discussion of OH masers at cosmological distances.
These produce errors of less than 10 percent in $d_l$ and less than 40
percent in comoving volume.

The incremental frequency band $d\nu$ observed at the frequency $\nu$
is related to increments in redshift for the OH and HI lines:
\begin{equation}
d\nu = - \frac{\nu^2}{\nu_{OH}}dz_{OH}= - \frac{\nu^2}{\nu_{HI}}dz_{HI} 
\end{equation}
implying that $dz_{OH}=(\nu_{OH}/\nu_{HI})dz_{HI}$. The flux $S$ measured
in the line emission from the OH megamaser of luminosity $L_{OH}$ is
falsely interpreted as HI emission from an object of $L_{HI}$:
\begin{equation}
S =  \frac{ L_{OH}}{4\pi(z_{OH}c/H_o)^2}=  \frac{ L_{HI}}{4\pi(zc/H_o)^2} 
\end{equation}
so that $L_{HI}=(z/z_{OH})^2L_{OH}$. Since the redshift range being 
explored for imitation HI signals is small, $z<0.01$, the OH masers
are drawn from a thin shell, allowing $z_{OH}$ to be considered
constant at $z_{OH}\approx 0.17$ in what follows.

The goal is to derive an approximation for the imitation HI mass
function $\phi_{im}$ provided by distant OH masers. 
This can be accomplished by
evaluating the number of signals that appear to come from the volume
$dV_{HI}$ from emitters of strength $L_{HI}$ but which were actually produced
in $dV_{OH}$ from emitters of $L_{OH}$.  The relation between the
luminosity functions then is 
$\Theta(L_{OH}) dL_{OH}dV_{OH}=\phi_{im}(L_{HI}) dL_{HI}dV_{HI}$.
Solving for $\phi_{im}$, substituting from the above expressions as
necessary and then including the approximate relation for the 
OH luminosity function from Eq.(\ref{approxtheta.eqn}), which is
valid here since all the OH signals detected will originate in
emitters much brighter than $L_{OH}^*$:
\begin{eqnarray}
\label{imhisteps.eqn}
\phi_{im} &=& \Theta(L_{OH})\left(\frac{dL_{OH}}{dL_{HI}}\right)
                \left(\frac{dV_{OH}}{dV_{HI}}\right) 
\\
&=& \Theta\left(\left(\frac{z_{OH}}{z}\right)^2L_{HI}\right)
                  \left(\frac{z_{OH}}{z}\right)^4 
                 \frac{\nu_{OH}}{\nu_{HI}}
\nonumber \\
&=& \frac{C(\alpha+\beta)f_m}{2L_{OH}^*}
     \frac{\nu_{OH}}{\nu_{HI}}
     \left(\frac{L_{HI}}{2L_{OH}^*}\right)^{-\frac{\alpha+\beta+2}{2}}
      \left(\frac{z_{OH}}{z}\right)^{2-\alpha-\beta}
\nonumber
\end{eqnarray}
This result can then be recast as a dependence on $M_{HI}$:
\begin{eqnarray}
\phi_{im}(M_{I})dM_{HI} &=& Cf_m\frac{\alpha+\beta}{2}
         \left(\frac{L_{OH}^*}{L_{HI}^*}\right)^{\frac{\alpha+\beta}{2}}
\nonumber \\
  & &  {\times}	\,\,\, \frac{\nu_{OH}}{\nu_{HI}}
\left(\frac{M_{HI}}{M_{HI}^*}\right)^{-\frac{\alpha+\beta+2}{2}}
\nonumber \\
  & &   {\times} \,\,  \left(\frac{z}{z_{OH}}\right)^{\alpha+\beta-2}
      \frac{dM_{HI}}{M_{HI}^*}
\label{imhifin.eqn}
\end{eqnarray}
After substitution for the constants in Eq.(\ref{imhifin.eqn}), $\phi_{im}$
has the functional dependence
\begin{eqnarray}
\phi_{im}(M_{HI})dM_{HI} &\propto&  
\left(\frac{M_{HI}}{M_{HI}^*}\right)^{-2.15}
\left(\frac{z}{z_{OH}}\right)^{0.3}
      \frac{dM_{HI}}{M_{HI}^*}
\label{imhiimp.eqn}
\end{eqnarray}
producing a form with that mimics a diverging low luminosity tail,
since $(\alpha+\beta+2)/2 = 2.15$ is greater than 2.  The dependence on
$z$ does not affect the power law shape of the imitation mass function but
does weakly affect the normalization that would be inferred.

The number of imitation HI masses per decade of $M_{HI}$ becomes
\begin{equation}
\hat\phi_{im}\approx 2.3{\times}10^{-5}
   \left(\frac{M_{HI}}{M_{HI}^*}\right)^{-1.15}
   \left(\frac{z}{z_{OH}}\right)^{0.3} {\rm Mpc}^{-3}
\end{equation}
This estimate implies that a survey sensitive to 1~mJy signals,
corresponding to $M_{HI}$ between $10^5$ and $10^6$M$_{\odot}$
at velocities of 300 to 500~km~s$^{-1}$, would be unlikely to 
detect more than one imitator per cubic megaparsec.  Therefore, with the current
set of parameters describing the OH luminosity function, significant
contamination of HI surveys at low redshift
would only occur if the HI mass function
turns out to be as flat as $\gamma\approx 1$ or drops off
at low $M_{HI}$. Even substantial evolution
in the merging rate from $z=0$ to 0.17 is unlikely to raise this
density estimate by more than a factor of two, but some caution is in order
until our understanding of the properties of OH masers improves.
The likelihood for confusion might rise significantly if surveys were
conducted at a 0.2 mJy sensitivity.

\section{Conclusion}
 
The density of OH megamaser galaxies in the sky may be high enough
that radio spectroscopic surveys will be effective tools at identifying
them.  The time evolution of the megamaser sources could then be monitored
by a series of surveys performed at a range of radio frequencies (i.e. 
different redshifts).

OH megamasers may form a source of confusion in surveys designed to
detect neutral hydrogen in normal galaxies through their 21 cm emission.
High resolution radio mapping, optical spectroscopy and far
infrared detection would remove the ambiguity.

\begin{acknowledgements} 
  The author is grateful to W.A. Baan,
  G.D. Bothun, A.G. de Bruyn, and J.M. van
  der Hulst for discussions and comments.
  This research has made use of the NASA/IPAC Extragalactic Database
  (NED) which is operated by the Jet Propulsion Laboratory, California
  Institute of Technology, under contract with the National
  Aeronautics and Space Administration.

\end{acknowledgements}

\end{document}